\documentclass[prc,aps,epsfig]{revtex4}

\usepackage{psfig}

\def\lsim{\raise0.3ex\hbox{$<$\kern-0.75em\raise-1.1ex\hbox{$\sim$}}}
\def\gsim{\raise0.3ex\hbox{$>$\kern-0.75em\raise-1.1ex\hbox{$\sim$}}}

\def\beq{\begin{equation}}
\def\eeq{\end{equation}}
\def\beqa{\begin{eqnarray}}
\def\eeqa{\end{eqnarray}}

\newcommand{\rr}{\mbox{\boldmath $r$}}
\newcommand{\rrn}{\mbox{$r$}}

\newcommand{\rb}{\mbox{\boldmath $b$}}

\def\gappeq{\mathrel{\rlap {\raise.5ex\hbox{$>$}}
{\lower.5ex\hbox{$\sim$}}}}

\def\lappeq{\mathrel{\rlap{\raise.5ex\hbox{$<$}}
{\lower.5ex\hbox{$\sim$}}}}

\def\Toprel#1\over#2{\mathrel{\mathop{#2}\limits^{#1}}}

\begin{document}

\title{Testing universality of the Color Glass Condensate descriptions} 
\author{  M.S.Kugeratski$^1$, V.P. Gon\c{c}alves$^2$, and  F.S. Navarra$^1$}
\affiliation{$^1$Instituto de F\'{\i}sica, Universidade de S\~{a}o Paulo, 
C.P. 66318,  05315-970 S\~{a}o Paulo, SP, Brazil\\
$^2$High and Medium Energy Group (GAME), \\
Instituto de F\'{\i}sica e Matem\'atica,  Universidade
Federal de Pelotas\\
Caixa Postal 354, CEP 96010-900, Pelotas, RS, Brazil\\}
\begin{abstract}

Perturbative Quantum Chromodynamics (pQCD) predicts that the small-$x$ gluons in a 
hadron wavefunction should form a Color Glass Condensate (CGC), which has universal 
properties, which are the same for all hadrons or nuclei.
Assuming this property, in this paper  we cross relate the current CGC descriptions 
of the $ep$ HERA data and $dAu$ RHIC data. In particular, we use  the quark dipole scattering amplitude 
recently proposed by Kharzeev, Kovchegov  and Tuchin (KKT) to explain 
the high $p_T$ particle suppression observed in $dAu$ collisions at RHIC in our 
calculations of the proton and 
longitudinal structure functions. We present a detailed comparison between this 
parameterization and those proposed to describe the $ep$ HERA data.  We find out that,
due to its  peculiar dependence on the  energy and dipole separation,  
the KKT parameterization is able to describe the experimental $ep$ data only in a limited kinematical range of photon virtualities.

\end{abstract}
\maketitle
\vspace{1cm}
\section{Introduction}

In  the past few years much theoretical
effort has been devoted towards the understanding of the growth of
the total scattering cross sections with energy. These studies are
mainly  motivated by the violation of the unitarity  (or
Froissart) bound by the solutions of the linear perturbative DGLAP \cite{DGLAP}
and BFKL \cite{BFKL} evolution equations. Since these evolution equations predict
that the cross section rises obeying a power law of the energy,
violating the Froissart bound \cite{FrMa}, new dynamical effects associated
with the unitarity corrections are expected to stop its further growth \cite{GLR,MUELLERS}. 
This expectation can be easily understood:
while for large momentum transfer $k_{\perp}$, the BFKL  equation
predicts that the mechanism $g \rightarrow gg$ populates the
transverse space with a large number of small size gluons per unit
of rapidity (the transverse size of a gluon with momentum
$k_{\perp}$ is proportional to $1/k_{\perp}$), for small
$k_{\perp}$ the produced gluons overlap and fusion processes, $gg
\rightarrow g$, are equally important. Considering the latter process,
the rise of the gluon distribution below a typical scale is reduced,
restoring the unitarity. That typical scale is energy dependent and is called  
saturation scale  $Q_s$.  The saturation momentum sets the
critical transverse size for the unitarization of the cross
sections. In other words, unitarity is restored by including
non-linear corrections in the evolution equations 
\cite{GLR,MUELLERS,BARTELS,AYALAS,VENUGOPALAN,BAL,CGC,WEIGERT,KOVCHEGOV,BRAUN}.
Such effects are small for $k_{\perp}^2 > Q_{\mathrm{s}}^2$ and very strong for $k_{\perp}^2
< Q_{\mathrm{s}}^2$, leading to the saturation of the scattering amplitude.

\begin{figure}[t] 
\vspace*{0.5cm}
\centerline{\psfig{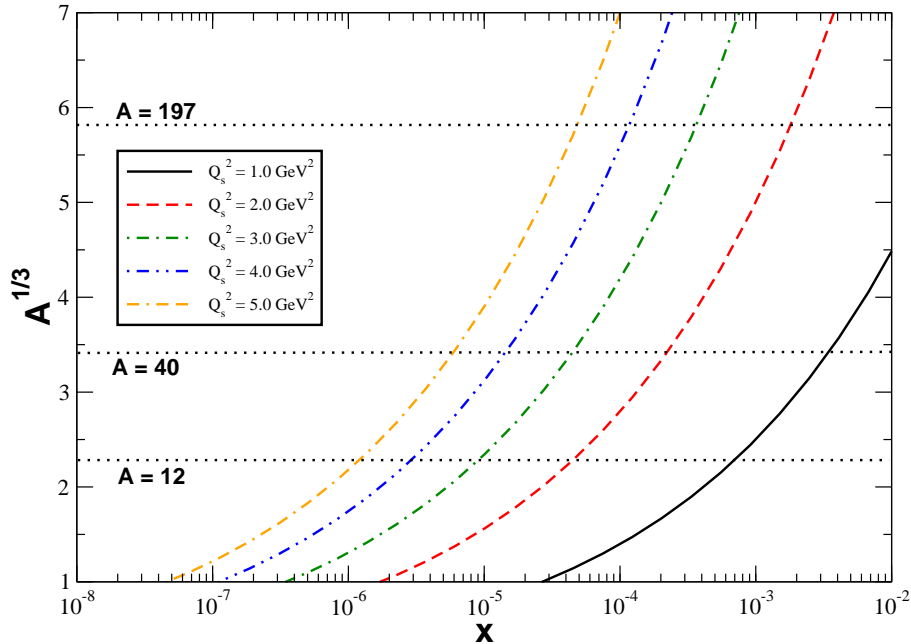}}
\caption{Saturation scale for different values of $A$ and $x$.}
\label{satscale}
\end{figure}

In the high energy limit, perturbative Quantum Chromodynamics (pQCD) predicts that the 
small-$x$ gluons in a hadron wavefunction should form a Color Glass Condensate (CGC), which is 
 described by an infinite hierarchy of the coupled evolution equations for the correlators of 
Wilson lines \cite{VENUGOPALAN,BAL,CGC,WEIGERT}.  In the absence of correlations, the first 
equation in the Balitsky-JIMWLK hierarchy decouples and is then equivalent to the equation 
derived independently by Kovchegov within the dipole formalism \cite{KOVCHEGOV}. 
A complete analytical solution of the Balitsky-Kovchegov (BK) equation is still lacking 
though there have been interesting recent developments in this direction (for recent 
reviews see, e.g. \cite{iancu_raju,anna_review,weigert_review,kov_jamal_review}). 
A remarkable feature which emerges from the solution of this equation is that the dense, 
saturated system of partons to be formed in hadronic wave functions at high energy has 
universal properties, the same for all hadrons or nuclei.  In particular, as the parton 
densities present  in dAu collisions at RHIC are not too different from those measured in 
DIS at HERA, one expects CGC physics (and thus the presence of an energy dependent saturation 
scale $Q_s$) to affect particle production rates and cross sections. This allows us to cross 
relate these experiments in this respect and gain a clear understading of the CGC in high 
energy experiments. In order to illustrate this statement, in Fig. \ref{satscale} we present 
the $A$ and $x$ dependence of the saturation scale, assuming the empirical parameterization 
$Q_s^2 = A^{\frac{1}{3}} \times Q_0^2 \, (\frac{x_0}{x})^{\lambda}$, with the parameters 
$Q_0^2 = 1.0$ GeV$^2$, $x_0 = 0.267 \times 10^{-4}$ and $\lambda = 0.253$ as in 
Ref. \cite{iancu_munier}. We can observe that, while in the proton case we need very 
small values of $x$ to obtain large values of $Q_s^2$, in the nuclear case a similar value 
can be obtained for values of $x$ approximately two orders of magnitude greater. In particular, 
the value of  $Q_s^2 = 2$ GeV$^2$, which is  estimated from $ep$ HERA data, can be obtained in 
$dAu$ collisions at RHIC  in the forward rapidity region. 
A strong support for the universality of the CGC physics has been given recently in 
Ref. \cite{armesto_prl}, which has noticed that the results for different collision 
systems in $\gamma^* \, p\, (A)$, $dA$ and $AA$ can be related through the geometric 
scaling property, which is one
of the main characteristics of the high density QCD approaches \cite{IANCUGEO,STASTOGEO}.

 The search of signatures for the parton saturation effects has been an
active subject of research in the last years (for recent reviews see, e.g. 
\cite{iancu_raju,kov_jamal_review,vicmag_mpla}).
In particular, it has  been observed that the HERA data at small $x$ and low $Q^2$ can be
successfully described with the help of saturation models \cite{GBW,bgbk,kowtea,iancu_munier,fs}. 
Moreover,  experimental results for the total cross section \cite{scaling} and also for   
inclusive charm production \cite{prl} present the property of  geometric scaling.  On the other hand, the recently  observed \cite{BRAHMSdata}
suppression of high $p_T$ hadron yields at forward rapidities in dAu collisions at RHIC 
has the behavior  anticipated on the basis of CGC ideas \cite{cronin}. Although the data is qualitatively 
consistent with the
predictions based on the CGC picture, only recently more quantitative analysis have been 
made \cite{jamal,kkt05} (See also Refs. \cite{armesto_prl,Dumitru}).  
These approaches consider different basic assumptions in order to describe the experimental 
data. In particular, they consider distinct prescriptions for the dipole target cross 
section which is one of the basic elements of the CGC approaches.  In Ref. \cite{jamal} 
a generalization of the parameterization proposed by Iancu, Itakura and Munier (IIM) to 
describe the HERA data was used, obtaining a good agreement with the BRAHMS data on charged hadron production in the limited region of low transverse momenta and  forward rapidity ($y = 3.2$). A comparison between this model and the RHIC data in the full kinematical range is not possible due to the behavior of the Fourier transform of the IIM dipole target cross section at intermediate transverse momenta \cite{jamalpri} (For a recent detailed discussion of this subject see Ref. \cite{betemps}). 
 In \cite{kkt05}, Kharzeev, Kovchegov and Tuchin (KKT)
introduced a new parameterization with the free parameters fitted to RHIC data.
In order to describe the hadron production in $d Au$ collisions at forward and mid-rapidities the authors has considered the contributions of gluon and valence quark production and convoluted it with the fragmentations functions and deuteron parton distributions. In particular, the gluon production cross section is given in terms of the gluon dipole scattering amplitude ${\cal{N}}_G(\rr,x)$, while the valence quark production cross section is a function of the quark dipole scattering amplitude ${\cal{N}}_Q(\rr,x)$. In principle both ${\cal{N}}_G(\rr,x)$ and ${\cal{N}}_Q(\rr,x)$ should be determined from the solution of the BK (or JIMWLK) evolution equation. However, as an analytical solution of this equation has not been accomplished so far, the authors from Ref. \cite{kkt05} have proposed a phenomenological parameterization for these two scattering amplitudes, inspired in the approximated analytical solutions of the BK equation for the saturation and color transparency regimes. It is important to emphasize that the pair separation and energy dependence  proposed for the scattering amplitudes ${\cal{N}}_G(\rr,x)$ and ${\cal{N}}_Q(\rr,x)$ are identical, characterized by a $(\rr  Q_s^2)^{\gamma(Y,\rr^2)}$ dependence, where the form of the anomalous dimension  
$\gamma(Y,\rr^2)$ is constructed considering known analytical solutions to the BFKL equation. As our goal in this paper is to analyze the saturation physics in the deep inelastic scattering processes, which is directly associated to the quark dipole scattering amplitude, we will only consider the expression for  ${\cal{N}}_Q(\rr,x)$ proposed in Ref.  \cite{kkt05}. The main uncertainty present in this procedure is associated with the normalization of the dipole cross section, which comes from the impact parameter dependence, and is not specified in  \cite{kkt05}. In what follows we will consider the normalization as a free parameter to be fixed in a comparison with the experimental data and keep all other parameters fixed as in Ref.  \cite{kkt05}.

 Based on 
the universality of the hadronic 
wave function,  we might expect that the KKT parameterization would also describe the HERA 
data on  proton structure functions. The main goal of this paper is to check this 
expectation. We will  compare the  predictions made with the KKT cross section  
and   HERA  $ep$ data in the  kinematical 
region where the saturation effects should be present (small $x$ and low $Q^2$). 
Moreover, we analyze in detail the quark dipole scattering amplitude proposed in Ref. \cite{kkt05} and 
compare with those previously proposed to describe  HERA data (for a related discussion 
see Ref. \cite{jamal_photons}). We observe  that there are large differences in the energy and 
pair dipole size dependences of these models. We will arrive at the conclusion that 
the experimental data on proton structure function are described using the  KKT prescription for the dipole cross section only in a limited kinematical range of photon virtualities. This result can be interpreted as an indicative that pre-asymptotic effects cannot be disregarded for the kinematical range of the RHIC and HERA colliders, which implies that the property of universality is still not manifested. It is important to emphasize that numerical studies of the BK equation show that its solution for intermediate rapidities presents a strong dependence in the choice for the initial condition \cite{BRAUN,solBK,lubli}. Other important aspect that deserves more detailed analyzes is that the impact parameter dependence, which is disregarded in the phenomenological parameterizations, may have a significant effect on the behavior of the dipole scattering amplitudes. 

A comment is in order here. A systematic comparison between the IIM and KKT dipole cross sections was started in 
Refs. \cite{jamal_photons} and \cite{kov_jamal_review}, where possible ways to choose which one is the most
appropriate were discussed. It was pointed out that one promising observable is the nuclear 
modification factor  for photon production in deuteron-gold collisions at $y=3.8$. In 
\cite{jamal_photons} it was  mentioned that the KKT parameterization had not been checked  
against DIS data on proton targets at HERA. In this paper we perform this check.

This paper is organized as follows. In the  next section we shortly review the deep inelastic 
scattering in the color dipole picture, where the relation between the proton structure 
function $F_2(x,Q^2)$ and the dipole target cross section $\sigma_{dip}$ becomes explicit. 
Moreover, we review the parameterizations proposed in the literature for the quark dipole scattering amplitude, 
with particular emphasis on that proposed in Ref. \cite{kkt05}. In Section \ref{results} 
we present a detailed comparison between the distinct parameterizations for ${\cal{N}}(\rr,x)$ 
discussed in the previous section.
 The asymptotic predictions for the color transparency and black disk regimes  are compared as 
well as the energy and dipole size dependences. Furthermore, a  comparison of   predictions 
with the $F_2$ HERA data in the kinematical region of small values of $x$ and $Q^2$ is 
presented. As a by product, we also present a comparison with the HERA data on the 
longitudinal structure function. Finally, in Section \ref{conc} we summarize our main 
results and conclusions.

\section{Deep Inelastic Scattering}

We start from the space-time picture of the eletron-proton/nuclei
processes \cite{dipole}. The deep inelastic scattering $ep(A)
\rightarrow e + X$ is characterized by a large electron energy
loss $\nu$ (in the target rest frame) and an invariant momentum
transfer $q^2 \equiv - Q^2$ between the incoming and outgoing
electron such that $x = Q^2/2m_N \nu$ is fixed ($m_N$ is the target mass). 
In terms of Fock
states we then view the $ep(A)$ scattering as follows: the
electron emits a photon ($|e\!> \rightarrow |e\gamma\!>$) with
$E_{\gamma} = \nu$ and $p_{t \, \gamma}^2 \approx Q^2$. Afterwards  the
photon splits into a $q \overline{q}$ ($|e\gamma\!> \rightarrow |e
q\overline{q}\!>$) and typically travels a distance $l_c \approx
1/m_N x$, referred to as the coherence length, before interacting in
the target. For small $x$, the photon is converted into  a quark
pair at a large distance before the scattering.
Consequently, the space-time picture of the DIS in the target rest
frame can be viewed as the decay of the virtual photon at high
energy  into a quark-antiquark pair (color dipole), which subsequently interacts with the 
target (for a review see, e.g., Ref. \cite{PREDAZZI}). In the small $x$ region, the color 
dipole crosses the
target with fixed transverse distance $\rr$ between the
quarks. The interaction $\gamma^*p(A)$ is further factorized and
is given by \cite{dipole},
\begin{eqnarray}
\sigma_{L,T}^{\gamma^*p(A)}(x,Q^2)= \sum_f \int dz \,d^2\rr
|\Psi_{L,T}^{(f)}(z,\rr,Q^2)|^2
\,\sigma_{dip}^{p(A)}(x,\rr),
\end{eqnarray}
where $z$ is the longitudinal momentum fraction of the quark of flavor $f$. The
photon wave functions $\Psi_{L,T}$ are determined from light cone
perturbation theory and are given by  
\begin{eqnarray}
|\Psi_{T} (z,\rr,\,Q^2)|^2  =  \frac{6\alpha_{\mathrm{em}}}{4\,\pi^2}  \, 
\sum_f e_f^2 
 \, {[z^2 + (1-z)^2]\, \varepsilon^2 \, K_1^2(\varepsilon \,\rrn)
 + m_f^2 \, \, K_0^2(\varepsilon\, \rrn)} & &
\label{wtrans}
\end{eqnarray}
and
\begin{eqnarray}
|\Psi_{L} (z,\rr,\,Q^2)|^2  =  \frac{6\alpha_{\mathrm{em}}}{\pi^2} \,
\sum_f e_f^2 
\, \left\{Q^2 \,z^2 (1-z)^2 \, K_0^2(\varepsilon\, \rrn) \right\} .  & &
\label{wlongs}
\end{eqnarray}
 The variable $\rr$ defines the relative transverse
separation of the pair (dipole) and $z$ $(1-z)$ is the
longitudinal momentum fraction of the quark (antiquark). The
auxiliary variable $\varepsilon^2=z(1-z)\,Q^2 + m^2_f$ depends on
the quark mass, $m_f$. The $K_{0,1}$ are the McDonald functions
and the summation is performed over the quark flavors.

The dipole hadron cross section $\sigma_{dip}$  contains all
information about the target and the strong interaction physics.
In the Color Glass Condensate (CGC)  formalism \cite{BAL,CGC,WEIGERT}, 
$\sigma_{dip}$ can be
computed in the eikonal approximation and is given by:
\begin{eqnarray}
\sigma_{dip} (x,\rr)=2 \int d^2 \rb \, {\cal{N}}(x,\rr,\rb)\,\,,
\end{eqnarray}
where ${\cal{N}}$ is the  quark dipole-target forward scattering amplitude for a given impact 
parameter $\rb$  which encodes all the
information about the hadronic scattering, and thus about the
non-linear and quantum effects in the hadron wave function. The
function ${\cal{N}}$ can be obtained by solving an appropriate evolution
equation in the rapidity $Y\equiv \ln (1/x)$. The main properties
of ${\cal{N}}$ are: (a) for the interaction of a small dipole ($\rr
\ll 1/Q_{\mathrm{s}}$), ${\cal{N}}(\rr) \approx \rr^2$, implying  that
this system is weakly interacting; (b) for a large dipole
($\rr \gg 1/Q_{\mathrm{s}}$), the system is strongly absorbed and therefore 
${\cal{N}}(\rr) \approx 1$.  This property is associated  to the
large density of saturated gluons in the hadron wave function.  
It is useful to assume that the impact parameter dependence of $\cal{N}$ can be factorized as 
${\cal{N}}(x,\rr,\rb) = {\cal{N}}(x,\rr) S(\rb)$, so that 
$\sigma_{dip}(x,\rr) = {\sigma_0} \,{\cal{N}}(x,\rr)$, with $\sigma_0$ being   a free 
parameter  related to the non-perturbative QCD physics.

Several models for the dipole cross section have been used in the literature in order to fit 
the HERA data. Here we will consider only the models proposed in Refs. \cite{GBW,iancu_munier} 
which capture the main properties of the CGC physics. An  
equally good fit has been obtained in Ref. \cite{lubli}, where the $x$ dependence of 
the dipole cross section was derived from the numerical solution of the BK equation,  
including DGLAP corrections.
In Ref. \cite{GBW}  Golec-Biernat and Wusthoff (GBW) have proposed a 
  phenomenological
saturation model where ${\cal{N}}$ is given by 
\begin{eqnarray}
{\cal{N}}(x,\rr)  =  
\left[\, 1- \exp \left(-\frac{\,({ Q_{\mathrm{s}}(x)\,\rr})^2}{4} \right) \, \right]
\label{n_gbw}
\end{eqnarray}
with $Q_s^2 = Q_0^2\,e^{\lambda\ln(x_0/x)}$. 
The parameters
were obtained from a fit to the HERA data yielding
$\sigma_0=23.03 \,(29.12)$ mb, $\lambda= 0.288 \, (0.277)$ and
$x_0=3.04 \cdot 10^{-4} \, (3.41 \cdot 10^{-4})$ for a 3-flavor
(4-flavor) analysis~\cite{GBW}.  An
additional parameter is the effective light quark mass, $m_f=0.14$
GeV, consistent with the pion mass. It should be noticed that the
quark mass plays the role of a regulator for the photoproduction
($Q^2=0$) cross section. The light  quark mass is one of the
non-perturbative inputs in the model. The charm quark mass is
considered to be $m_c=1.5$ GeV. A smooth transition to the
photoproduction limit is obtained with a modification of the
Bjorken variable as,
\begin{eqnarray}
\tilde{x}= x\, \left( \, 1+ \frac{4\,m_f^2}{Q^2}
\,\right)=\frac{Q^2 + 4\,m_f^2}{W^2} \,.
\end{eqnarray}
Observing Eq. (\ref{n_gbw}) we notice that when
$Q_s^2(x)\,\rr^2\ll 1$, the model reduces to
color transparency, whereas as  one approaches the region
$Q_s^2(x)\,\rr^2 \approx 1$, the exponential
takes care of resumming many gluon exchanges, in a
Glauber-inspired way. Intuitively, this is what happens when the
proton starts to look dark. 
Although the GBW parameterization gives a good description of the old HERA data, it has 
been ruled out by the new HERA data, with a much higher accuracy. This shortcoming is mainly 
related to the fact that this model fails to describe the Bjorken scaling violation and
its functional form is only an approximation of the theoretical non-linear QCD approaches.

Another CGC inpired model has been proposed to described the HERA data in Ref. 
\cite{iancu_munier}. It is based on the 
 understanding of the BFKL approach in the border of the saturation region \cite{IANCUGEO}. 
In particular, the forward scattering amplitude has been calculated in both leading order (LO)  and next-to-leading order (NLO) BFKL  
approaches in the geometric scaling region \cite{BFKLSCAL}. It reads 
\begin{eqnarray}
{\cal{N}}(x,\rr)=\,\left[\rr^2 Q_s^2(x)\right]^{\gamma_s}\,
\exp\left[ -\frac{\ln^2\,\left(\rr^2 Q_s^2\right)}{2\,\beta \,\bar{\alpha}_sY}\right]\,,
\label{sigmabfkl}
\end{eqnarray}
where the power $\gamma_s$ is the (BFKL) saddle point in the vicinity of the saturation 
line $Q^2= Q_{\mathrm{s}}^2(x)$.
In this model the overall normalization of the dipole cross section is given by 
 $\sigma_0  =2\pi R_p^2$, where $R_p$ is the proton radius.
 In addition, the  anomalous dimension is defined as $\gamma = 1- \gamma_s$. As usual in 
the BFKL formalism, $\bar{\alpha}_s=N_c\,\alpha_s/\pi$ and  $\beta \simeq 28\,\zeta (3)$. 
The quadratic diffusion factor in the exponential gives rise to the scaling violations, 
which are essential to describe the HERA data.
As the forward scattering amplitude  in Eq. (\ref{sigmabfkl}) does not include an 
extrapolation from the geometric scaling region to the saturation region, the authors 
from Ref. \cite{iancu_munier} have constructed  a parameterization for  
${\cal{N}} (x,\rr)$ which smoothly interpolates between the  limiting behaviors 
analytically under control: the solution of the BFKL equation
for small dipole sizes, $\rr\ll 1/Q_s(x)$, and the Levin-Tuchin law \cite{Levin}
for larger ones, $\rr\gg 1/Q_s(x)$. A fit to the structure function $F_2(x,Q^2)$ 
was performed in the kinematical range of interest, showing that it is  not very 
sensitive to the details of the interpolation (for a comprehensive phenomenological 
analysis of the HERA results using the numerical solution of the BK equation see 
Ref. \cite{lubli}). The  dipole-target forward scattering amplitude  was parametrized 
as follows,
\begin{eqnarray}
{\cal{N}}(x,\rr) =  \left\{ \begin{array}{ll} 
{\mathcal N}_0\, \left(\frac{\rr\, Q_s}{2}\right)^{2\left(\gamma_s + 
\frac{\ln (2/\rr Q_s)}{\kappa \,\lambda \,Y}\right)}\,, & \mbox{for $\rr Q_s(x) \le 2$}\,,\\
 1 - \exp^{-a\,\ln^2\,(b\,\rr\, Q_s)}\,,  & \mbox{for $\rr Q_s(x)  > 2$}\,, 
\end{array} \right.
\label{CGCfit}
\end{eqnarray}
where the expression for $\rr Q_s(x)  > 2$  (saturation region)   has the correct 
functional
form, as obtained either by solving the Balitsky-Kovchegov (BK) equation 
\cite{BAL,KOVCHEGOV}, 
or from the theory of the Color Glass Condensate (CGC) \cite{iancu_raju}. Hereafter, 
we label the model above by IIM. The coefficients $a$ and $b$ are determined from the 
continuity conditions of the dipole cross section  at $\rr Q_s(x)=2$. The coefficients 
$\gamma_s= 0.63$ and $\kappa= 9.9$  are fixed from their LO BFKL values. In our 
further calculations we shall  use the parameters $R_p=0.641$ fm, $\lambda=0.253$, 
$x_0=0.267\times 10^{-4}$ and ${\mathcal N}_0=0.7$, which give the best fit result. 
Recently, this model has also been used in phenomenological studies of  vector meson 
production \cite{Forshaw1} and  diffractive processes \cite{Forshaw2} at HERA as well 
as for the description of the longitudinal structure function \cite{vicmag_fl}. 

On the other hand, Kharzeev, Kovchegov and Tuchin (KKT) have proposed a new 
parameterization for the dipole scattering amplitude in order to describe  hadron 
production in dAu collisions  \cite{kkt05}. 
As already discussed in the Introduction, in order describe the hadron production in $d Au$ collisions at forward and mid-rapidities these authors 
have proposed a phenomenological parameterization for the quark and gluon dipole  scattering amplitudes, inspired in the approximated analytical solutions of the BK equation for the saturation and color transparency regimes. 
In this model the expression 
for the quark dipole-target forward scattering amplitude (hereafter ${\cal{N}}_Q= {\cal{N}}(\rr,x)$)  is given by
 \cite{kkt05}:
\beq\label{glauber2}
{\cal{N}}(\rr,x) \, = \, 1- \exp\left[-\frac{1}{4} \left(\rr^2
\frac{C_F}{N_c} \, Q_s^2\right)^{\gamma(Y,\rr^2)}\right].
\eeq
where   the  anomalous dimension $\gamma(Y, \rr^2)$  is
\beq\label{gamma}
\gamma(Y, \rr^2) \, = \, \frac{1}{2}\left(1+\frac{\xi 
(Y, \rr^2)}{\xi (Y,\rr^2) + \sqrt{2 \,\xi (Y, \rr^2)}+  7
\zeta(3)\, c} \right),
\eeq
where $c$ is a free parameter and 
\beq\label{xi}
\xi (Y, \rr^2) \, = \, \frac{\ln\left[1/( \rr^2 \, Q_{s0}^2 ) 
\right]}{(\lambda/2)(Y-Y_0)}\,.
\eeq
The authors assume that the saturation scale can be expressed by $
Q_s^2(Y)  = \Lambda^2 A^{1/3} \left(\frac{1}{x}\right)^{\lambda}$. 
 The form of the anomalous
dimension is inspired by the analytical solutions to the BFKL equation
\cite{BFKL}. Namely, in the limit $\rr\rightarrow 0$ with $Y$ fixed we 
recover the anomalous dimension in the double logarithmic
approximation $\gamma \approx 1 - \sqrt{1/(2 \, \xi)}$. In another
limit of large $Y$ with $\rr$ fixed, Eq. (\ref{gamma}) reduces to the
expression of the anomalous dimension near the saddle point in the
leading logarithmic approximation $\gamma \approx
\frac{1}{2} + \frac{\xi}{14 \, c \, \zeta (3)}$. Therefore Eq. (\ref{gamma}) 
mimicks the onset of the geometric scaling region \cite{iancu_munier,IANCUGEO}. 
In the calculations of  Ref. \cite{kkt05} it is assumed that a 
 characteristic value of $\rr$ is $\rr
\approx 1/(2 \, k_T)$ where $k_T$ is the transverse momentum of the valence 
quark and $\gamma$ was approximated by 
$\gamma(Y, \rr^2) \approx \gamma(Y,1/(4 \, k_T^2))$. As our goal is to apply this 
model to deep inelastic scattering,  we explore two other   
possible approximations  which are $\rr \approx 1/ Q_s$ and $\rr \approx 1/ Q$. 
In the above expressions the parameters $\Lambda=0.6$~GeV and
$\lambda=0.3$ are fixed by DIS data \cite{GBW}. The initial saturation scale used
in (\ref{xi}) is defined by $Q_{s0}^2=Q_s^2(Y_0)$ with $Y_0$ being  the
lowest value of rapidity at which the low-$x$ quantum evolution
effects are essential. When applied to describe  RHIC data,  the amplitude 
${\cal{N}}(\rr,x)$ must be convoluted with the quark distribution function in the
hadron and with the fragmentation function of the quark. Moreover the  
gluon contribution must be added. These procedures introduce  uncertainties in the 
predictions, which can only be estimated if the formalism is applied to other processes.

A comment is in order here. 
 The main goal of the  IIM and KKT parameterizations is to mimick  CGC physics  
in all kinematical regions. 
However, currently we  have some theoretical control only on the asymptotic regimes 
of saturation and color transparency. Therefore we must  assume some 
interpolation Ansatz in order to obtain a parameterization that may be used in practical 
calculations. Moreover,  although both parameterizations have similar form,  the KKT 
parameterization, in contrast to IIM one, includes the double logarithmic limit as 
well as the correct behavior in the saturation and geometric scaling regions. As is 
well known, this limit is important for large transverse momentum, allowing to 
connect the CGC physics with the DGLAP predictions. However, what is the correct 
linear limit (BFKL/DGLAP) in the kinematical regions of  HERA and RHIC is still an 
open question.

\section{Results and discussion}
\label{results}

\begin{table}[t]
\begin{center}
\begin{tabular} {||c|c|c||}
\hline
\hline
    &  $\rr \ll \rr_{sat}$  &  $\rr \gg \rr_{sat}$  \\
\hline
\hline
 GBW  & $ \frac{(\rr Q_s)^2}{4}$  & 1 \\
\hline
IIM & $  { {\cal N}_0} \, \left(\frac{{ \rr Q_s}}{2}\right)^{2\left({ \gamma_s} + 
\frac{\ln(2/{ \rr Q_s})}{ \kappa { \lambda} Y}\right)}$  &  $ 1 - {e}^{- a\ln^2(b\,  
{ \rr Q_s})} $\\
\hline
KKT & $ \frac{1}{4} \left(\frac{C_F}{N_c}\, ({ \rr
    \,Q_s})^2\right)^{\gamma(Y,\rr^2)}$  &  1 \\

\hline
\hline
\end{tabular}
\end{center}
\caption{Asymptotic limits of the quark dipole scattering amplitude  in  
different models.}
\label{tab1}
\end{table}

\begin{figure}[t] 
\vspace*{0.5cm}
\centerline{\psfig{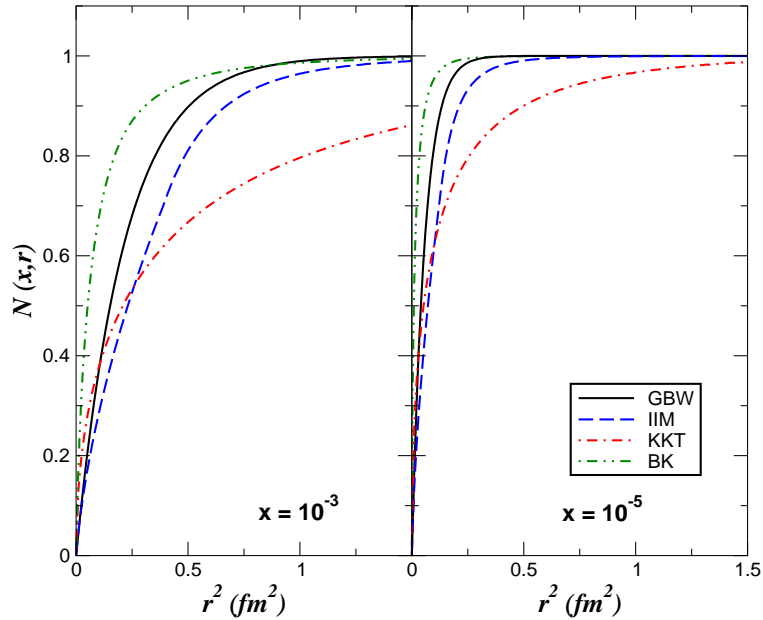}}
\caption{Dependence of the quark dipole scattering amplitude  in the squared pair separation $\rr^2$ at 
different values of $x$.}
\label{size}
\end{figure}

In this section we present a detailed study between the predictions of the distinct 
parameterizations for the quark dipole scattering amplitude  and a comparison of its predictions with HERA data.  
We start presenting in Table \ref{tab1} the
asymptotic predictions for the linear regime $\rr \ll \rr_{sat}$ and saturation regimes  
$\rr \gg \rr_{sat}$, where $\rr_{sat} \equiv 1/Q_s$.
As discussed in the Introduction, the critical line dividing dense and dilute regions 
is the saturation scale $Q_s$, with the property that the smaller the $x$, the denser  
the  system gets and partons start to reinteract.
The basic feature of the GBW, IIM and KKT models is that for 
a given $\rr$, these models predict   that the amplitudes tend to unity at small 
values of $x$ in contrast to the linear solution which predicts a exponential growth 
in this kinematical region. Moreover, all these models predict that the system saturates 
early, that is for large values of $x$  when the dipole size is larger. 
The  three parameterizations present similar functional forms for the forward 
scattering amplitude in the two limits, with the IIM presenting a residual  $\rr Q_s$ 
dependence in the saturation regime, but also showing  saturation  for large values of $\rr Q_s$.

In Fig.\ref{size} we analyze  the pair separation dependence of 
the quark dipole scattering amplitude for different values of $x$. 
As expected from the previous discussion, we observe that while the GBW and IIM 
parameterizations present a similar behavior for small $\rr^2$, the KKT one predicts 
a smoother dependence. In the other limit, the  GBW and IIM parameterizations saturate 
for large pair separations,  while the KKT one still presents a residual dependence, 
demonstrating that the asymptotic regime is only reached for very large pair separations. 
The characteristic feature which is evident in the GBW and IIM models is that the dipole 
cross section saturates for smaller dipoles when $x$ assumes  smaller values.
An important aspect to be emphasized is the large difference between the predictions in 
the transition region, which we expect to be probed at  HERA. 
For comparison we also present the predictions for ${\cal{N}}$  from the numerical solution of the BK equation as obtained in Ref. \cite{lubli}.  We have that this solution has, as expected, the color transparency and saturation limits for small and large $\rr$, respectively. However, the transition region is characterized by a sharp transition in the two values of $x$ shown. Moreover, it is important to emphasize the large difference between this results and the KKT prediction. In what follows we will restrict our analyzes for the phenomenological parameterizations which has its parameters fixed by the experimental data.

\begin{figure}[t]
\vspace*{0.5cm}
\centerline{\psfig{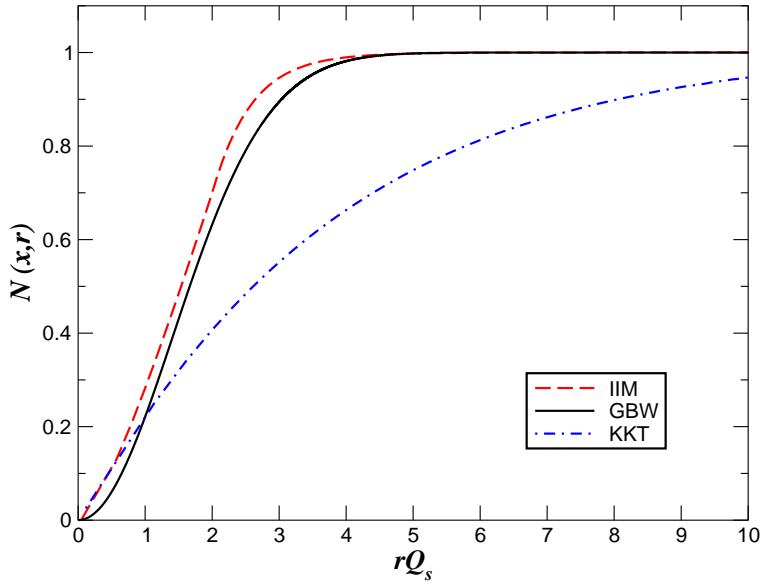}}
\caption{Quark dipole scattering amplitude as a function of the scaling variable $\rr Q_s$.}
\label{rqs}
\end{figure}

\begin{figure}[t] 
\vspace*{0.5cm}
\centerline{\psfig{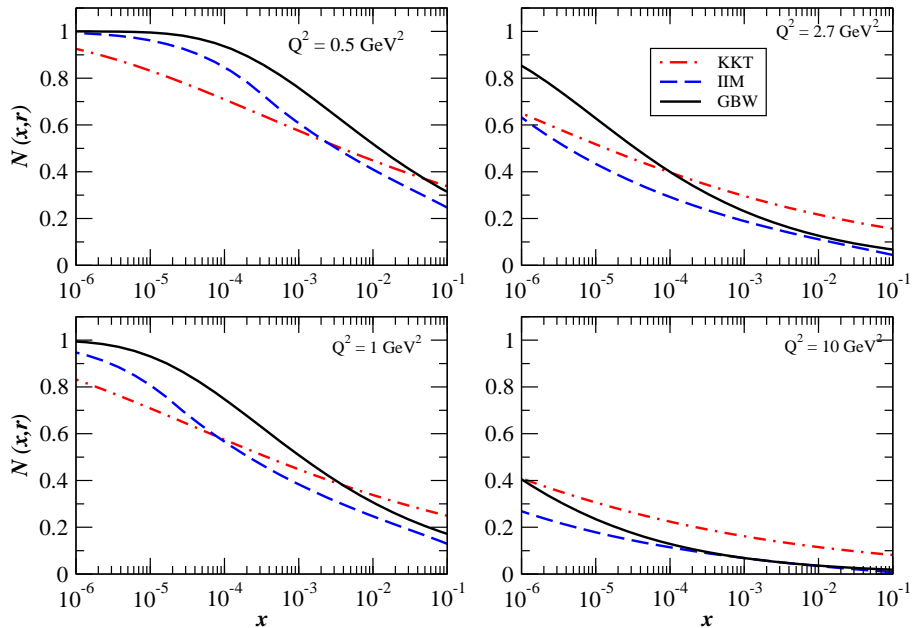}}
\caption{Energy dependence of the quark dipole scattering amplitude for different values of the 
squared pair separation $\rr^2 = 1/Q^2$. }
\label{energy}
\end{figure}

All models have the property of geometric 
scaling  observed in the solutions of the BK equation. Mathematically, geometrical 
scaling means that the solution of the BK equation depends only on one combined 
variable $\rr Q_s(x)$ instead of $\rr$ and $x$ separately, i. e., 
${\cal{N}}(\rr,x) \equiv {\cal{N}}(\rr Q_s(x))$.
In  Fig. \ref{rqs} we show the dependence of the quark dipole scattering amplitude on the scaling 
variable $\rr Q_s$. We observe that the three dipole scattering amplitudes  grow in the region of small values of 
$\rr Q_s$ as a power of $\rr Q_s$, i.e. $ {\cal{N}}(\rr,x) \propto  
(\rr Q_s)^{2 \gamma_{eff}}$. However,   $\gamma_{eff}$ is different in 
each model, being 1 for the GBW model, $\le 1$ for the IIM model and about 
$\frac{1}{2}$ for the KKT one. This implies a different $\rr Q_s$ dependence 
of the dipole scattering amplitudes and dipole cross sections.  Since the saturation scale  drives 
the energy dependence of the dipole cross section, these models 
present a very distinct energy dependence. This can clearly be seen  in Fig. \ref{energy}, 
where we present the $x$ dependence of the dipole scattering amplitudes  for different values of 
the squared pair separation given by $\rr^2 = 1/Q^2$. 
We observe that for large $Q^2$ (small pair separation) the dipole scattering amplitude  
is dominated by the linear limit. Since the models have different behavior in this 
limit, the energy dependence is also different, with the GBW model 
presenting the strongest growth at small $x$. The behavior predicted by the IIM model is 
similar to the GBW one. On the other hand, the KKT model predicts 
the smallest growth with the energy. 
At large pair separations $\rr > \rr_s$, which characterizes the saturation regime,   
the GBW and IIM models predict the saturation of the dipole scattering ampitude, while the KKT one still presents a growth at small values $x$.  Basically, 
the asymptotic saturation regime is only observed for very small values of $x$, 
beyond the kinematical range of HERA.

\begin{figure}
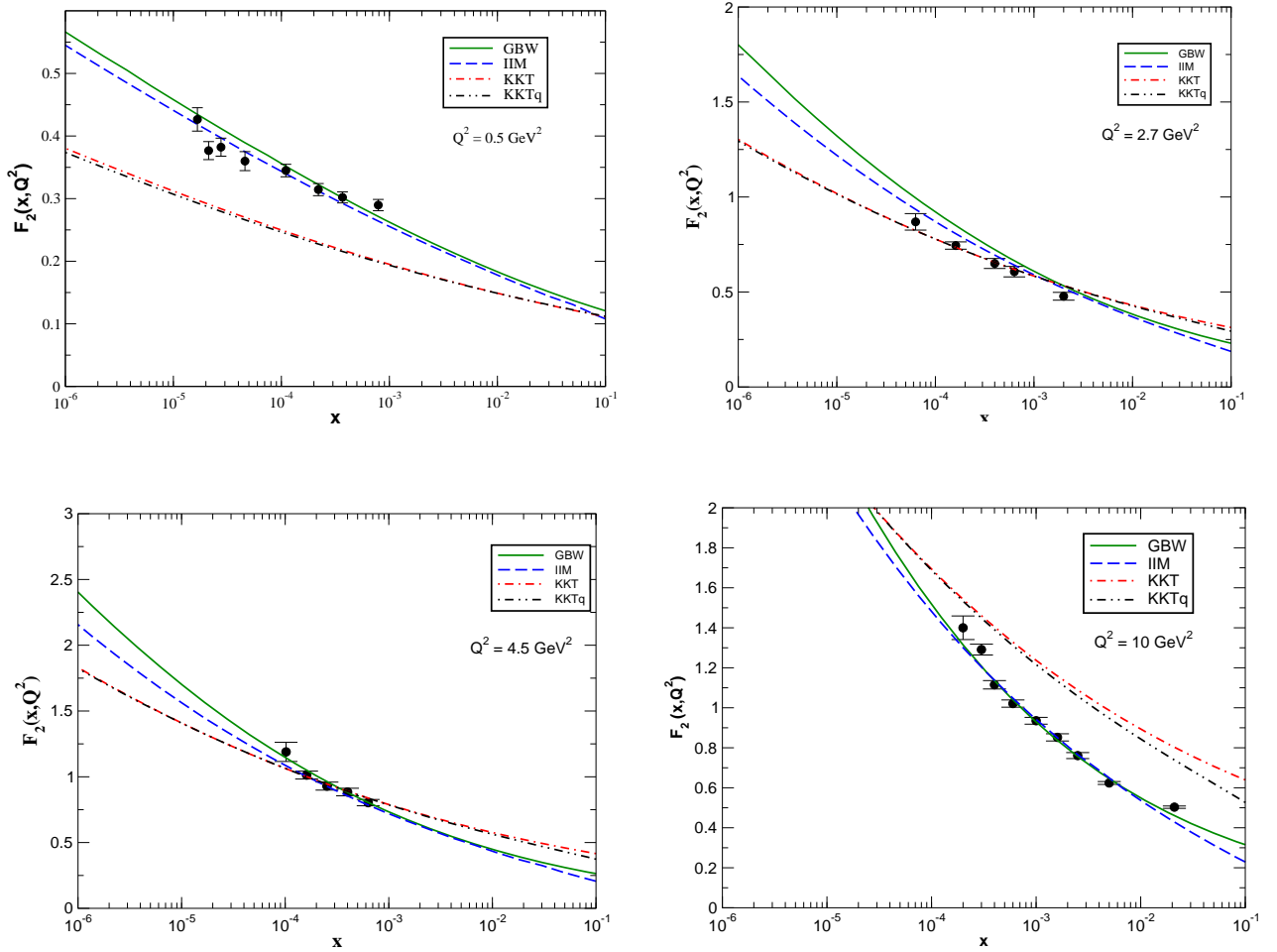

\centerline{
\begin{tabular}{ccc}
{\psfig{figure=f2q2_0_5b.eps,width=8.2cm}} & \,\,\,\,\,\, &
{\psfig{figure=f2q2_2_7b.eps,width=7.6cm}}\\
& & \\
& & \\
{\psfig{figure=f2q2_4_5b.eps,width=8.0cm}} & \,\,\,\,\,\, &
{\psfig{figure=f2q2_10b.eps,width=8.0cm}}
\end{tabular}}
\caption{Comparison between the predictions for $F_2(x,Q^2)$ of the distinct models at 
different values of $Q^2$. Data are from ZEUS.}
\label{fig1f2}
\end{figure}

\begin{figure} 
\vspace*{0.5cm}
\centerline{\psfig{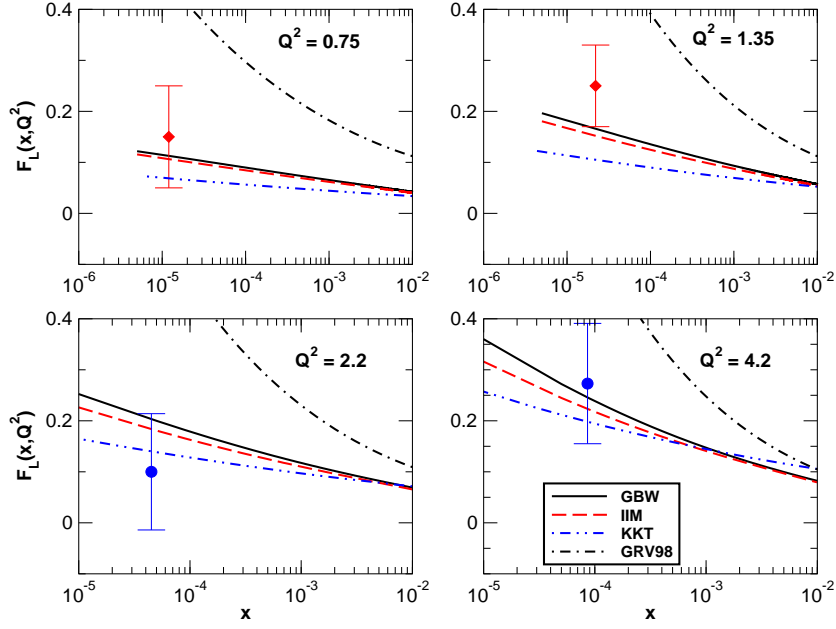}}
\caption{Comparison between the predictions for $F_L(x,Q^2)$ of the distinct models at  
different values of $Q^2$.}
\label{fig2f2}
\end{figure}

The basic observable  measured with a great accuracy by HERA is the proton structure 
function $F_2(x,Q^2)$ which is directly related with the $\gamma^* p$ cross section by 
the following expression
\beq
F_2(x, Q^2) = \frac{Q^2}{4 \pi^2 \alpha_{em}} (\sigma_T^{\gamma^* p} 
+ \sigma_L^{\gamma^* p}) \,\,.
\label{struc}
\eeq
Consequently, using the color dipole picture of DIS we can directly calculate 
$F_2$ for the  different parameterizations of the quark dipole scattering ampitudes.
Similarly, we can estimate the longitudinal structure function which is defined 
by $F_L(x, Q^2) = Q^2/(4 \pi^2 \alpha_{em}) \times  \sigma_L^{\gamma^* p}$.
In Figs. \ref{fig1f2} and \ref{fig2f2} we present a comparison between the predictions 
of the different models and  ZEUS data \cite{zeus}. 
 We have used that $\sigma_{dip}(x,\rr) =
 {\sigma_0} \,{\cal{N}}(x,\rr)$ with $\sigma_0$ as given in the GBW and IIM 
parameterizations. For the KKT parameterization we have treated $\sigma_0$ as 
a free parameter and fixed its value by fitting  the $F_2$ data at $Q^2 = 2.7$ GeV$^2$. Our choice for this value of virtuality is justified by the fact that in this region we expect that the saturation physics should be dominant. We have tested other choices and verified that our main conclusion is not modified (see below). The  predicitions for other values of virtualities are parameter free. Moreover,   
 we have considered two different choices for the typical 
scale present in the process, needed  to calculate the function $\gamma(Y, \rr^2)$ in the KKT parameterization. 
Basically, we have assumed that  $\rr \approx 1/ Q_s$ or $\rr \approx 1/ Q$. As we will 
demonstrate below, our predictions for $F_2$ in the kinematical range of interest are 
almost identical.
We consider only few values of the photon virtuality  in the region of low $Q^2$, where 
the saturation effects must be important. As expected, the GBW and IIM models describe 
quite well the experimental $F_2$ data (See Fig. \ref{fig1f2}) . On the other hand, the KKT parameterization is 
able to describe 
the experimental $ep$ data only in a limited kinematical range of photon virtualities around of the virtuality where the normalization is fixed. 
The basic aspect of this parameterization is that the $Q^2$ dependence of the proton structure function cannot be described. Furthermore,  this parameterization predicts 
an  energy dependence, which is 
smoother  than observed in the data. This behavior is directly related 
to the behavior present in the dipole scattering amplitude. The curve denoted KKTq in the figure 
represents the results obtained assuming $\rr \approx 1/ Q$, while in the KKT curve we 
assume $\rr \approx 1/ Q_s$ in the calculation of $\gamma(Y, \rr^2)$. These 
two prescriptions differ appreciably only in the large $x$ and/or $Q^2$ region. 

Finally, in Fig. \ref{fig2f2} we present the predictions of the different models for the 
longitudinal structure function. For comparison, we also present the prediction obtained 
using the Altarelli-Martinelli equation and the GRV98 parameterization for the solution 
of the DGLAP evolution equation (for details see Ref. \cite{vicmag_fl}).  
In this case we have that the KKT parameterization describe  reasonably the few data available \cite{fldata}, similarly to the GBW and IIM parameterizations. However, this fact is mainly associated to the large experimental error in the current data. We believe that a future experimental study of the longitudinal structure function will be able to discriminate  the parameterizations.

\section{Summary}
\label{conc}
 Assuming the universality of the hadron wavefunction predicted by the Color Glass 
Condensate formalism we can cross relate different  experiments  and gain a clear 
understading of the QCD dynamics at  high energies. In this paper we have studied 
in detail  different parameterizations of the dipole scattering ampitude  proposed to describe 
the HERA and RHIC data. We have observed that these parameterizations predict distinct 
energy and dipole size dependences, mainly in the interpolation region between the linear 
and saturation regimes. Since the experimental data at HERA probe exactly this kinematical 
domain, a comparison with the $F_2$ data in the region of small $x$ and low $Q^2$ is very 
important, since it allows  to discriminate between the parameterizations. 
We have concluded that the KKT parameterization is 
able to describe 
the experimental $ep$ data only in a limited kinematical range of photon virtualities. Therefore, the  scaling violations of the proton structure function, observed in the   HERA data, are not reproduced by this model. Moreover, the KKT parameterization predicts  
a smoother energy 
dependence than that observed in the data.  As the IIM parameterization is not able to describe the RHIC data in the full kinematical range, our result put in check the property of universality, present in the CGC physics, for the current kinematical range of the RHIC and HERA experiments. In principle, it indicate that pre-asymptotic effects as for instance those associated to the different initial conditions present in $ep$ and $pA$ collisions, cannot still be disregarded and the cross relation between different experiments should be made with some caution as well as the interpretation of the comparison between the CGC predictions and the experimental data. We believe that an unified  global fit of the RHIC and HERA data could be useful to obtain reliable predictions for the future colliders.




\begin{acknowledgments}
The authors thank M. V. T.  Machado and M. A. Betemps for helpful discussions. This work was  partially 
financed by the Brazilian funding
agencies CNPq, FAPESP and FAPERGS.
\end{acknowledgments}


\end{document}